\documentclass[preprint,showkeys,aps,amssymb,prd]{revtex4}

\usepackage{bm,graphicx, wrapfig}
\usepackage{amsmath,amssymb,latexsym}

\newcommand\beq{\begin{equation}}
\newcommand\beqa{\begin{eqnarray}}
\newcommand\beqan{\begin{eqnarray*}}
\newcommand\eeq{\end{equation}}
\newcommand\eeqa{\end{eqnarray}}
\newcommand\eeqan{\end{eqnarray*}}

\newcommand\mdot{{\sf m}_\bullet}

\newcommand\alphahat{\hat{\alpha}}

\begin{document}

\title{Light's Bending Angle in the Equatorial Plane of a Kerr Black Hole} 

\author{S. V.  Iyer}
\email{iyer@geneseo.edu}
\author{E. C.  Hansen}
\email{ech3@geneseo.edu}
\affiliation{Department of Physics \& Astronomy, State University of New York at Geneseo,\\
1 College Circle,
Geneseo, NY 14454.}

\begin{abstract}
We present here a detailed derivation of an explicit spin-dependent expression for 
the bending angle of light as it traverses in the equatorial 
plane of a spinning black hole.  We show that the deflection produced in the 
presence of the black hole angular momentum explicitly depends  
on whether the motion of the light ray is in the direction, or opposite to the spin.  Compared to the zero-spin Schwarzschild
case, the bending angle is greater for direct orbits, and {\em smaller} 
for retrograde orbits, confirming our physical intuition about the loss of left-right symmetry from a
lensing perspective.  In addition, we show that for higher
spins, the effect is more pronounced resulting in tighter winding of direct orbits with respect to the
axis of rotation, and a higher degree of unwinding of retro orbits.  A direct 
consequence of this effect is a shift in image positions in strong gravitational lensing.

\end{abstract}

\keywords{gravitational lensing, bending angle, Kerr black holes}

\maketitle

\section{Introduction}
\label{sec:intro}

The deflection of light as it traverses through curved spacetime remains one of the 
few available observational tools to quantitatively study the geometry surrounding 
a strong source of gravitation.  The study of gravity's effect on light, or gravitational 
lensing, began with Einstein's prediction in 1913, and the subsequent confirmation 
by Eddington in 1919, of the bending angle of light in the relatively weak gravitational field near the sun.  The need 
for detailed study of gravitational lensing in both the weak and strong deflection regimes has become 
more compelling as the precision of our observational tools have grown many-fold in the past few decades.  

Bending angle calculations\cite{darwin, atkinson, luminet, ohanian} for Schwarzschild and Kerr geometries 
show that as we approach the depths of the gravitational potential, the bending angle exceeds $2\pi$ 
indicating that multiple looping of a light ray around the
center of attraction is possible (see for example, \cite{MTW73}, page 678).  In
strong deflection gravitational lensing, images formed as a result of this are referred
to as relativistic images.  In order to study these images, we need analytical expressions
for light deflection in exact form and sometimes as perturbative series expansions.
In this paper, we present a detailed analysis of deflection of light in the equatorial 
plane of the spinning, or Kerr, black hole.  Indeed, the program was 
started by Darwin\cite{darwin}, continued by Boyer and Lindquist\cite{boy-lind},
Chandrasekhar\cite{chandra} and many others since.  The key difference between our approach 
and earlier work is that we obtain an explicit expression for the bending 
angle for {\em both} direct and retrograde motion.  Our
final result for the bending angle is expressed 
explicitly in terms of the black hole mass $\mdot=GM/c^2$ and 
its spin parameter $a=J/Mc$, where $J/M$ is the 
angular momentum per unit mass of the black hole.  

Our goal is to continue to work in the same perturbative framework that was started in
\cite{iyerpetters} for the Schwarzschild black hole and apply it to the case 
of deflection of rays confined to the Kerr equatorial plane.
Bending angles, and therefore the position and magnification of images, 
depend crucially on whether the light ray is traversing in 
the same or opposite direction to the direction of rotation.
In the Schwarzschild case \cite{iyerpetters}, we expressed the bending angle
in terms of the invariant impact parameter $b$, and in fact found it useful there to use the 
variable $b'=1-b_c/b$, where $b_c=3\sqrt{3}\mdot$ was the Schwarzschild 
critical impact parameter.  For the equatorial Kerr case, we introduce a new definition for $b'$ that 
includes the non-zero spin of the black hole.  We are able to define precisely the approach
towards the critical impact parameter which is referred to as the strong deflection limit (SDL).  
The weak deflection limit (WDL), on the other hand, in both the Schwarzschild and the Kerr case are 
easily defined in terms of the impact parameter as the limit $b\rightarrow\infty$.  The detailed
SDL and WDL series are presented elsewhere \cite{iyerhansen2}.

In Section 2, we set up the basic framework with definitions of the variables 
and sign convention.  We also discuss the meaning of a number of 
limiting values as we cross-check our results with the zero-spin (Schwarzschild) case.  We 
obtain a formal exact expression for the bending angle in terms of elliptic integrals of the 
third kind.  A plot of the bending angle as a function of the impact parameter from the critical value all
the way to infinity is presented.  We believe that this plot has not appeared in literature before.

The exact bending angle is plotted numerically for both cases to show a remarkable difference 
between the deflection of light on the direct side and the retro side.  On the retro side, as the light ray traverses 
``upstream" it suffers a {\em smaller} deflection angle; smaller even than the Schwarzschild bending angle.  This 
unravelling of the light ray as it tries to loop around against the spin direction supports our physical intuition.  On the
direct side, the bending angle is greater than the Schwarzschild value, resulting in a tighter looping of 
the light ray.  Moreover, we show that as we vary the spin parameter $a$ from low to high
values, relativistic images are shifted inward on the direct side, and outward on the retro 
side.  We derive series expansions of the bending angle in 
the strong and weak deflection limits, to be presented elsewhere \cite{iyerhansen2}, and show 
that the analytical results can be used to predict image positions and magnifications.  

\section{The Exact Bending Angle for Kerr Geometry}
\label{sec:exactkerr}

We begin with the Kerr line element expressed in the Boyer-Lindquist
coordinates (with $\theta=\pi/2$ for the equatorial plane):

\beq \label{eq-Kerr-metric1}
  {\rm d}s^2 = g_{tt}{\rm d}t^2 
    \ + \ g_{rr}{\rm d}r^2 \ + \ g_{\phi \phi}\,
    {\rm d}\phi^2 \ + \ 2 g_{t \phi}\, {\rm d}t {\rm d}\phi
\eeq
where
\beqa
g_{tt}(r)=-\left(1-\frac{2\mdot}{r}\right)\\
g_{rr}(r)=\left(1-\frac{2\mdot}{r}+\frac{a^2}{r^2}\right)^{-1}\\
g_{\phi\phi}(r)=r^2+a^2+\frac{2\mdot a^2}{r}\\
g_{t\phi}(r)=\frac{-2\mdot a}{r}
\eeqa
where $t = c \tau$ and $\mdot=G M/c^2$ (gravitational radius) 
with $\tau$ physical time and $M$ the physical mass of the black hole.  From the 
Euler-Lagrange equations for null geodesics, we have
\beq
{\dot{r}}^2=\left[{\cal E}^2+\frac{{\cal E}^2a^2}{r^2}
+\frac{2M{\cal E}^2a^2}{r^3}-\frac{4Ma{\cal E}J_z}{r^3}
-\frac{J_z^2}{r^2}+\frac{2MJ_z^2}{r^3}\right]\nonumber
\eeq
and
\beq
\dot{\phi}=\frac{J_z\left[1-\dfrac{2\mdot}{r}+\dfrac{2\mdot a}{r}\dfrac{\cal E}{J_z}\right]}{(r^2-2\mdot r+a^2)}
\eeq

\begin{figure}[htbp] 
\begin{center} 
\includegraphics[width=3in]{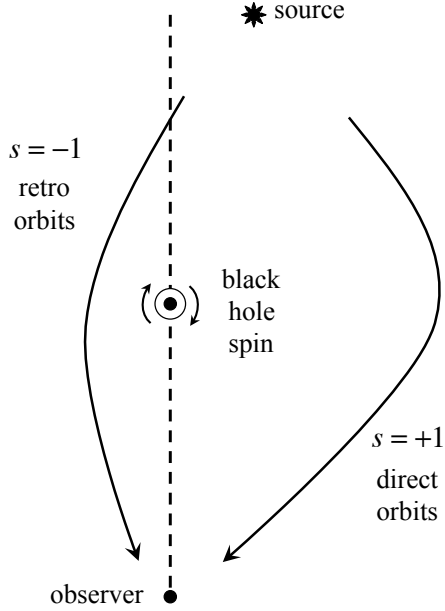}
\caption{\small \sl Sign convention for orbits as viewed from above.  The spin axis points into
the page in this figure.} 
\label{OrbitSign} 
\end{center} 
\end{figure}

Couched in the quantity $J_z$ that appears even at this early stage, is the sign of the 
angular momentum.  This will decide whether the light ray is traversing along the
direction of frame-dragging or opposite to it.  We will use the following notation to keep
track of the two types of orbits as we go through the analysis.  We define
\beq
b_s=s\left|\frac{J_z}{{\cal E}}\right|\equiv s b
\eeq
with $s={\rm Sign}(J_z/{\cal E})$ and $b$ is the positive magnitude.  The parameter $s$
is positive for direct orbits and negative for retrograde orbits as shown in Figure \ref{OrbitSign}.

With $a=J/Mc$ it is convenient to introduce the following notation:
\beq
\hat{a}=\frac{a}{\mdot}=\frac{Jc}{GM^2}
\eeq
limiting ourselves to cases where $0\le \hat{a} \le 1$, with $\hat{a}=0$ being the Schwarzschild limit 
and $\hat{a}=1$ being extreme Kerr.  Orbits with $s=+1$ and $b_s>0$ will be referred to as 
direct or prograde orbits; and those with $s=-1$ and $b_s< 0$
as retrograde orbits.  

Rewriting the above expression, we have
\beq
{\dot{r}}^2=J_z^2\left[\frac{1}{b^2}
+\frac{a^2}{b^2r^2}+\frac{2\mdot a^2}{r^3 b^2}
-\frac{4\mdot a}{r^3 b_s}-\frac{1}{r^2}+\frac{2\mdot}{r^3}\right]\nonumber
\eeq
Next, letting $u=1/r$ and using
\beq
\left(\frac{du}{d\phi}\right)^2
=\frac{u^4 (\dot{r})^2}{(\dot{\phi})^2},
\eeq
we have
\beq
(\dot{\phi})^2=J_z^2\left[\frac{1-2\mdot u
+2\mdot au/b_s}{a^2-(2\mdot/u)+(1/u^2)}\right]^2
\eeq
and
\beq
(\dot{r})^2=J_z^2\left[\frac{1}{b^2}
+2\mdot u^3 \left(1-\frac{a}{b_s}\right)^2
-u^2\left(1-\frac{a^2}{b^2}\right)\right].\nonumber
\eeq
Combining the above expressions and simplifying gives us
\beq \label{polynomials}
\left(\frac{du}{d\phi}\right)^2
=\frac{\left[1-2\mdot u+a^2 u^2\right]^2}
{\left[1-2\mdot u\left(1-a/b_s\right)\right]^2} B(u),
\eeq
where the quantity $B(u)$, a cubic polynomial given by
\beq \label{cubic}
B(u)=\left[2\mdot \left(1-\frac{a}{b_s}\right)^2 u^3
-\left(1-\frac{a^2}{b^2}\right)u^2+\frac{1}{b^2}\right]
\eeq
has a maximum of two positive roots and 
at most one negative root.  We consider the case of one 
real negative root $u_1$ and two real distinct 
positive roots $u_2$ and $u_3$.  The three roots, given 
in terms of two intermediate constants $P$ and $Q$ that 
allow us to line up the roots in the order $u_1<u_2<u_3$ are given by 
\beqa
u_1 &=&\frac {P-2\mdot-Q}{4\mdot r_0} \\
\nonumber\\
u_2 &=&\frac{1}{r_0} \\
\nonumber\\
u_3 &=&\frac{P-2\mdot+Q}{4\mdot r_0}
\eeqa

By comparing the coefficients in $B(u)$ to those 
in the original polynomial in equation(\ref{cubic}) we 
first obtain the following relationship 
between $P$ and $\{a, b, s, r_0\}$
\beq
P=r_0 \frac{\left(1-\dfrac{a^2}{b^2}\right)}
{\left(1+\dfrac{a}{b_s}\right)^2} \ 
= \ r_0 \frac{\left(1+\dfrac{a}{b_s}\right)}
{\left(1-\dfrac{a}{b_s}\right)},
\eeq
where $a$ and $b$ are positive quantities.  This gives us the following
relation between the critical parameters:
\beq
r_{sc}=3 \mdot \frac{\left(1-\dfrac{a}{b_{sc}}\right)}
{\left(1+\dfrac{a}{b_{sc}}\right)}.
\eeq
Note that we need to set $r_0=r_{sc}$, $b_s=b_{sc}$ and $P=3\mdot$ to 
obtain the critical values.  Note that when $a=0$, 
$r_{sc}\rightarrow r_c=3\mdot$ in both cases.  The critical value for $r_0$ is less than $3\mdot$ for 
direct orbits and greater than $3\mdot$ for retro orbits.

Comparing the other coefficients of the cubic 
polynomial $B(u)$ we obtain the following additional expressions: 
\beqa
\frac{Q^2-(P-2\mdot)^2}{8\mdot r_0^3}
=\frac{1}{b^2 \left(1-a/b_s\right)^2}\\
 \nonumber \\
Q^2=(P-2\mdot) (P+6\mdot).
\eeqa
The intermediate variables $Q$ and $P$ can be eliminated by combining 
the above relations to yield a simple cubic equation involving
the impact parameter and the distance of closest approach given by
\beq
r_0^3-b^2\frac{{\left(1-\dfrac{a}{b_s}\right)}^3}
{\left(1+\dfrac{a}{b_s}\right)} r_0 +2\mdot b^2
{\left(1-\dfrac{a}{b_s}\right)}^2=0.\nonumber
\eeq
which can be solved to yield
\begin{widetext}
\beq
r_0=\frac{2b}{\sqrt{3}}\sqrt{1-\dfrac{a^2}{b^2}}\;
\cos{\left[\frac{1}{3}\cos^{-1}{\left(\frac{-3\sqrt{3}\mdot}{b}\frac{\left(1-\dfrac{a}{b_s}\right)^2}
{{\left(1-\dfrac{a^2}{b^2}\right)}^{3/2}}\right)}\right]}
\eeq
\end{widetext}
This expression for the distance of closest approach and the invariant impact 
parameter becomes important in both the SDL and WDL series expansions \cite{iyerhansen2} completely in
terms of the invariant normalized quantity $b'$.  Also, note that 
this expression reduces to its Schwarzschild version (see e.g., equation (9) in \cite{iyerpetters}) when we 
set the spin parameter to zero.  

So, in the strong deflection limit, we have $P=3\mdot$ and
the following expressions involving the critical quantities:
\beq \label{rcrit}
r_{sc}=3 \mdot \frac{\left(1-\dfrac{a}{b_{sc}}\right)}
{\left(1+\dfrac{a}{b_{sc}}\right)}
\eeq
and 
\beq \label{bcrit}
{(b_{sc}+a)}^3=27 \mdot^2 (b_{sc}-a).
\eeq
Equations (\ref{rcrit}) and (\ref{bcrit}) agree 
with Chandrasekhar (eqn 82, p329).  

Combining these two equations gives us the following relationship 
between the critical values
\beq
b_{sc}^2=3 r_{sc}^2+a^2
\eeq
To solve the cubic equation (\ref{bcrit}) we need to 
consider the direct and retrograde motion separately.  For 
direct orbits (i.e., $s=+1$) we have
\beq
{(b_{+c}+a)}^3-27 \mdot^2 (b_{+c}+a)+54 a \mdot^2=0
\eeq
the solution of which is given by
\beq
b_{+c}=-a+6\mdot\cos{\left[\frac{1}{3}
\cos^{-1}{\left(\frac{-a}{\mdot}\right)}\right]}.
\eeq

\begin{figure}[htbp] 
\begin{center} 
\includegraphics[width=3.3in]{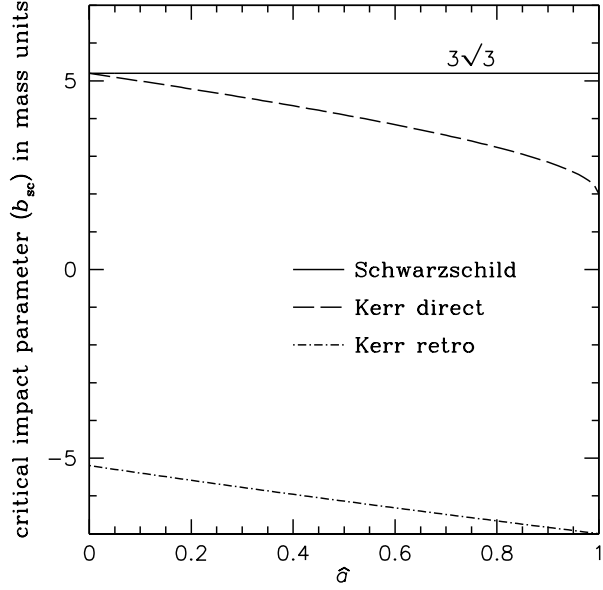}
\caption{\small \sl The critical impact parameter in units of $\mdot$ as a function
of the spin parameter $\hat{a}$.  The Schwarzschild critical impact parameter
is shown for reference.} 
\label{bcritvSpin} 
\end{center} 
\end{figure} 

\begin{figure}[htbp] 
\begin{center} 
\includegraphics[width=3.3in]{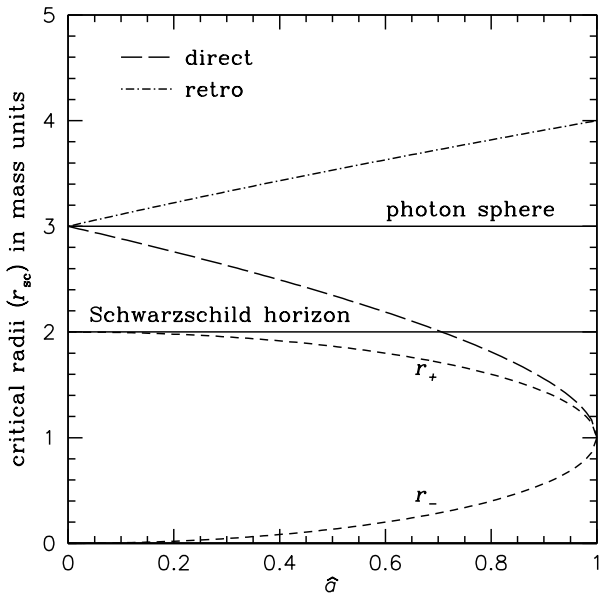}
\caption{\small \sl The critical radius in units of $\mdot$ plotted as a function
of the spin parameter $\hat{a}$.  The Schwarzschild radius, the radius of the 
photon sphere and the Kerr inner and outer horizon radii are shown for reference.} 
\label{rcritvSpin} 
\end{center} 
\end{figure}

For retrograde orbits with $s=-1$ we have
\beq
{(b_{-c}-a)}^3-27 \mdot^2 (b_{-c}-a)-54 a \mdot^2=0.
\eeq
Solving this cubic equation yields
\beq
b_{-c}=-a-6\mdot\cos{\left[\frac{1}{3}
\cos^{-1}{\left(\frac{a}{\mdot}\right)}\right]}.
\eeq
We can combine the two cases and write
\beq
b_{sc}=-a+s6\mdot\cos{\left[\frac{1}{3}
\cos^{-1}{\left(\frac{-sa}{\mdot}\right)}\right]}
\eeq
Note that the critical impact parameter is a function of the black hole spin.  In Figure \ref{bcritvSpin}, we have
plotted the critical impact parameter as a function of $a$.  
From equation (\ref{bcrit}), we have the following:
\beqan
\frac{(b_{sc}-a)}{(b_{sc}+a)}&=& \frac{(b_{sc}+a)^2}{27 \mdot^2}\\
&=& \frac{1}{27\mdot^2}\left\{s6\mdot\cos\left[\frac{1}{3}\cos^{-1}
\left(\frac{-sa}{\mdot}\right)\right]\right\}^2\\
&=& \frac{36}{27}\cos^2\left[\frac{1}{3}\cos^{-1}
\left(\frac{-sa}{\mdot}\right)\right]
\eeqan
Using the above, we obtain the following expression for the critical 
value of the distance of closest approach:
\beq
r_{sc}=2\mdot\left\{1+\cos\left[\frac{2}{3}\cos^{-1}
\left(\frac{-sa}{\mdot}\right)\right]\right\}
\eeq
which agrees with equation 87 on page 330 in \cite{chandra}.  Figure \ref{rcritvSpin} is a plot of 
$r_{sc}/\mdot$ as a function of spin.  For $0\le a\le\mdot$, the critical value 
of $r_0$ lies in the following ranges: $\mdot\le r_{sc}\le3 \mdot$ for
direct orbits and $3\mdot\ge r_{sc}\ge 4\mdot$ for retro orbits.  Note that $r_{sc}$ stays 
positive throughout, given that this is the radial coordinate. 

In the extreme case when $a=M$, we 
have the following results:
\beqan
\text{direct:} \qquad
s &=& +1 \qquad r_{+c}
=\mdot \qquad b_{+c}=2\mdot  \\
\text{retrograde:} \qquad
s &=& -1 \qquad r_{-c}
= 4\mdot \qquad b_{-c}=-7\mdot
\eeqan
in agreement with equation 89 on page 330 in \cite{chandra}.

The sign of the impact parameter carries the information about whether the light
propagation is in the same or opposite direction to the spin.  This appears to be the most
important difference in our approach, one that yields results pertaining to retrograde motion 
in a natural way.  We believe that details involving retrograde motion are harder to extract when the 
sign of $a$ is used to keep track of the change in the direction of the black hole spin.  

There are two other polynomials that appear in 
equation (\ref{polynomials}).  We will denote the 
roots of $1-2\mdot u+a^2 u^2=0$ as $u_{\pm}$, given by
\beq
u_{\pm}=\frac{\mdot\pm\sqrt{\mdot^2-a^2}}{a^2}
\eeq
Using this notation we can now write the ratio of 
the polynomials in terms of partial fractions as
\beq
\frac{1-2\mdot u(1-\omega_s)}{1-2\mdot u+a^2u^2}
=\frac{C_+}{(u_+ -u)}+\frac{C_-}{(u_- -u)}
\eeq
Solving for $C_+$ and $C_-$, we obtain
\beq
C_+=\frac{(\mdot+\sqrt{\mdot^2-a^2})2\mdot 
\left(1-\omega_s\right)-a^2}{2a^2\sqrt{\mdot^2-a^2}}
\eeq
and
\beq
C_-=\frac{a^2-(\mdot-\sqrt{\mdot^2-a^2})2\mdot 
\left(1-\omega_s\right)}{2a^2\sqrt{\mdot^2-a^2}}
\eeq

Let us now consider a light ray that starts in the 
asymptotic region and approaches the black hole, with $r_0$ as the distance of closest 
approach.  It then emerges and reaches an observer who is also in an asymptotic 
region.  Although the change in the $\phi$ coordinate of the light ray appears in the 
equations of motion, we will consider the deflection of the light ray from its original 
path $\alphahat$.  The change in $\phi$ and the bending angle $\alphahat$
are simply related by a difference of $\pi$.  The bending 
angle can be written down in a straightforward manner 
by integrating the equation of motion for $\phi$ and subtracting
$\pi$ from it.  The integrals therein can be separated into 
two parts as shown below to make the lower limit equal 
to the smallest root $u_1$. The integrals are then readily expressed in terms of 
elliptic integrals of the third kind (see \cite{byrd}, for example) to 
obtain the exact expression for the bending angle.  These steps
are shown below.  

\begin{widetext}
\beqa
\alphahat &=& -\pi+\;\sqrt{\frac{2}{\mdot}}\;\;
\frac{C_+}{1-\omega_s}\int_0^{1/r_0} \frac{du}{(u_+ -u)
\sqrt{(u-u_1)(u-u_2)(u-u_3)}} \nonumber \\
\nonumber\\
&&\qquad \qquad\qquad+\; \sqrt{\frac{2}{\mdot}}\;\;\frac{C_-}{1-\omega_s}
\int_0^{1/r_0} \frac{du}{(u_- -u)\sqrt{(u-u_1)(u-u_2)(u-u_3)}}\nonumber\\
&=&-\pi+\;\sqrt{\frac{2}{\mdot}}\;
\frac{C_+}{1-\omega_s}\left[\int_{u_1}^{u_2}\frac{du}
{(u_+ -u)\sqrt{(u-u_1)(u_2-u)(u_3-u)}} \right. \nonumber \\
&&\qquad\qquad\qquad\qquad\qquad\qquad-\left. \int_{u_1}^0 \frac{du}{(u_+ -u)
\sqrt{(u-u_1)(u_2-u)(u_3-u)}}\right] \nonumber \\ 
&&\qquad+\;\sqrt{\frac{2}{\mdot}}\;\frac{C_-}{1-\omega_s} 
\left[ \int_{u_1}^{u_2}\frac{du}{(u_- -u)
\sqrt{(u-u_1)(u_2-u)(u_3-u)}}\right. \nonumber \\
&&\qquad\qquad\qquad\qquad\qquad\qquad\qquad-\left. \int_{u_1}^0 \frac{du}{(u_- -u)
\sqrt{(u-u_1)(u_2-u)(u_3-u)}}\right]\nonumber
\eeqa

\beq
\label{ExactKAlpha}
\alphahat=-\pi+\frac{4}{1-\omega_s}\sqrt{\frac{r_0}{Q}} 
\bigg\lbrace\Omega_+ \Bigl[\Pi(n_+,k)
-\Pi(n_+,\psi,k)\Bigr]+\Omega_- \Bigl[\Pi(n_-,k)-\Pi(n_-,\psi,k)
\Bigr]\bigg\rbrace, 
\eeq
\end{widetext}
where $\Pi(n_\pm,k)$ and $\Pi(n_\pm,\psi,k)$ 
are the complete and the incomplete elliptic integrals 
of the third kind respectively.  The argument $k^2$ is defined through the elliptic
integral as usual in the range $0\le k^2 \le 1$.  Note that in some references the variable 
is referred to as $k^2$ and in others simply as $k$.  The order in which the
arguments appear in $\Pi(n,\psi,k)$ also varies between different references 
and in Mathematica.

\noindent {\it Remark:} In Mathematica, the built-in mathematical
function for the incomplete elliptic integral of the third kind 
${\rm EllipticPi}[{\sf n},\phi,{\sf m}]$ is defined by
\beq
\int_0^\phi
\left[1 - {\sf n}\sin^2 \theta\right]^{-1}\left[1 - {\sf m} \sin^2 \theta\right]^{-1/2}d \theta \nonumber
\eeq
and the complete elliptic integral of the third kind is $ {\rm EllipticPi}[{\sf n, m}] = {\rm EllipticPi}[{\sf n},\pi/2, {\sf m}]$.
 
The other variables in the above expression are defined
as follows:
\beqa
\Omega_\pm&=&\frac{C_\pm}{u_\pm-u_1}\nonumber \\
k^2&=&\frac{Q-P+6\mdot}{2Q} \nonumber \\
\psi&=& \arcsin{\sqrt{\frac{Q+2\mdot-P}{Q+6\mdot-P}}} \nonumber \\
n_\pm&=&\frac{u_2-u_1}{u_\pm -u_1} \nonumber
\eeqa

It can be easily shown that in the limiting case 
when $a \rightarrow 0$, we have $\Omega_+ =1$, 
$\Omega_-=0$ and $n_+=0$ to give 
\beqa
\alphahat&=&-\pi+4\sqrt{\frac{r_0}{Q}}\left[\Pi(0,k)
-\Pi(\psi,0,k)\right] \nonumber \\
&=& -\pi+4\sqrt{\frac{r_0}{Q}}\left[K(k)-F(\psi,k)\right],
\eeqa
where $K(k)$ and $F(\psi,k)$ are the complete and 
incomplete elliptic integrals of the first kind 
respectively.  In addition, in the limit when $\mdot=0$ (i.e., $h\rightarrow 0$) we recover 
zero deflection as expected.

We will use the following convenient notation:
\beq
h=\frac{\mdot}{r_0}\qquad\omega_s=\frac{a}{b_s} \qquad
{\rm and}\qquad\omega_0=\frac{a^2}{\mdot^2}
\eeq
with $\omega_s$ taking on the appropriate sign for direct and 
retrograde orbit.  Note that in the limit \{$\omega_s, \omega_0 \rightarrow 0$\}, we
recover the zero-spin Schwarzschild case, and in the limit $h\rightarrow 0$, we have the 
zero-deflection flat metric limit.   Further, we define critical 
parameters analogous to the Schwarzschild case in \cite{iyerpetters}:
\beq
h_{sc}=\frac{1+\omega_s}{1-\omega_s}  \qquad
{\rm and}\qquad r_{sc}=\frac{3\mdot}{h_{sc}}
\eeq
We also define the variable
\beq
h'=1-\frac{3h}{h_{sc}}\equiv1-3\left(\frac{\mdot}{r_0}\right)\left(\frac{1-\omega_s}{1+\omega_s}\right)
\eeq
with
\beq
1-3\left(\frac{\mdot}{r_0}\right)\left(\frac{1-\omega_s}{1+\omega_s}\right)\quad 
\xrightarrow{\quad a \rightarrow 0\quad} \quad1-\frac{3\mdot}{r_0}. \nonumber
\eeq
We have introduced these different quantities for the Kerr case, keeping in mind that they should go over 
to those defined in the Schwarzschild case smoothly when $a$ is set equal to zero.  So, as shown above, 
as $a\rightarrow 0$, $h_{sc}\rightarrow 1$ and we recover the definition of $h'$ in \cite{iyerpetters}.  In both
cases, $h\rightarrow 0$ at critical, and $h\rightarrow 1$ as $r_0$ approaches infinity.

From a lensing perspective, we are interested in impact 
parameters beyond the critical value (SDL) extending all the way to infinity (WDL).  We define the dimensionless 
quantity $b'$ as 
\beq 
b'=1-\frac{s b_{sc}}{b}
\eeq
where the insertion of the quantity $s$ guarantees that the $b'$ stays between $0$ and $1$.  Note that 
this definition goes over naturally in the Schwarzschild limit:
\beq
1-\frac{s b_{sc}}{b}\quad \xrightarrow{\quad a \rightarrow 0\quad} \quad1-\frac{3\sqrt{3}\mdot}{b}\nonumber
\eeq

Now, some of the intermediate variables can be eliminated to rewrite all quantities in terms of 
$h, h_{sc},\omega_0$ and $\omega_s$ as follows:
\beq
\label{bigcoeff}
\frac{r_0}{Q}=\frac{1}{h_{sc}\sqrt{\left(1-\dfrac{2h}{h_{sc}}\right)
\left(1+\dfrac{6h}{h_{sc}}\right)}}
\eeq
\beq
k^2=\frac{\sqrt{\left(1-\dfrac{2h}{h_{sc}}\right)
\left(1+\dfrac{6h}{h_{sc}}\right)}
+\dfrac{6h}{h_{sc}}-1}{2\sqrt{\left(1-\dfrac{2h}{h_{sc}}\right
)\left(1+\dfrac{6h}{h_{sc}}\right)}}
\eeq
\beq
\psi=\arcsin{\sqrt{\frac{1-\dfrac{2h}{h_{sc}}-\sqrt{\left(1-\dfrac{2h}{h_{sc}}\right)
\left(1+\dfrac{6h}{h_{sc}}\right)}}{1-\dfrac{6h}{h_{sc}}
-\sqrt{\left(1-\dfrac{2h}{h_{sc}}\right)\left(1+\dfrac{6h}{h_{sc}}\right)}}}}
\eeq
\begin{widetext}
\beq
\Omega_\pm=\frac{\pm (1\pm\sqrt{1-\omega_0})
(1-\omega_s)\mp\omega_0/2}{\sqrt{1-\omega_0}\left( 1\pm\sqrt{1-\omega_0}
-\dfrac{\omega_0 h_{sc}}{4}\left[1-\dfrac{2h}{h_{sc}}-\sqrt{\left(1-\dfrac{2h}{h_{sc}}\right)
\left(1+\dfrac{6h}{h_{sc}}\right)}\;\right]\right)}
\eeq
\beq
\label{nplusminus}
n_\pm=\frac{1-\dfrac{6h}{h_{sc}}-\sqrt{\left(1-\dfrac{2h}{h_{sc}}\right)
\left(1+\dfrac{6h}{h_{sc}}\right)}}
{1-\dfrac{2h}{h_{sc}}-\sqrt{\left(1-\dfrac{2h}{h_{sc}}\right)
\left(1+\dfrac{6h}{h_{sc}}\right)}-\dfrac{4}{\omega_0 h_{sc}}
\left(1\pm\sqrt{1-\omega_0}\right)}
\eeq
\end{widetext}
All of the above variables are to be substituted into the bending angle expression.  We note here that the 
quantities $r_0,h, h_{sc}$, and $\omega_s$ depend on $b$, while $\omega_0=a^2/\mdot^2$ is 
independent of $b$.  Any quantity that has an ``$s$" in the subscript takes on a negative sign 
for retro orbits. The bending angle itself stays positive since the
sign of $\phi$ in the equations of motion is determined by the incident ray in the
asymptotic region.  In other words, as the ray approaches critical on the retro side, the overall deflection
is still towards the black hole even though the extent to which it is bent is smaller than in the static case.  
Another way of looking at this is to change the direction of the rotation axis from ``into the page" in 
Figure \ref{OrbitSign} to ``out of page" and look at the ray on the right side of the figure for which 
$s=-1$ now; the bending angle is still positive.

\begin{figure}[htp] 
\begin{center} 
\includegraphics[width=3.3in]{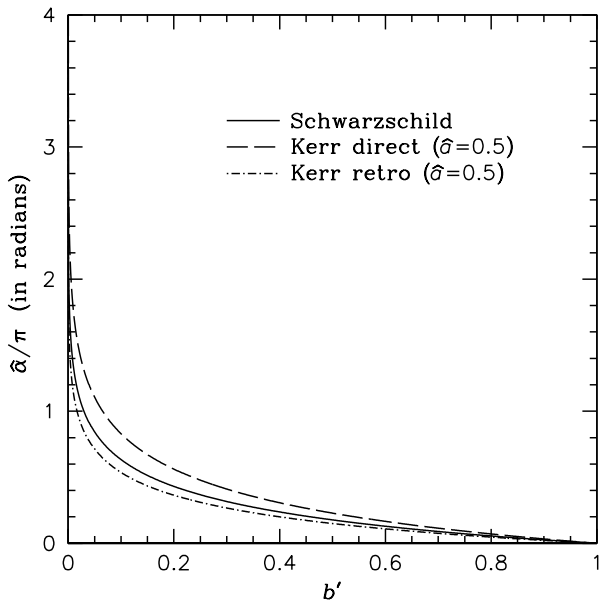}
\caption{\small \sl Exact bending angle plotted as a function of the normalized impact parameter.  The strong
deflection regime is towards the left of this plot as $b'$ goes to zero, and the weak deflection
regime is as $b'\rightarrow \infty$.  The Schwarzschild bending angle
is shown for reference.} 
\label{ExactKerrp5} 
\end{center} 
\end{figure}

\begin{figure}[htp] 
\begin{center} 
\includegraphics[width=3.3in]{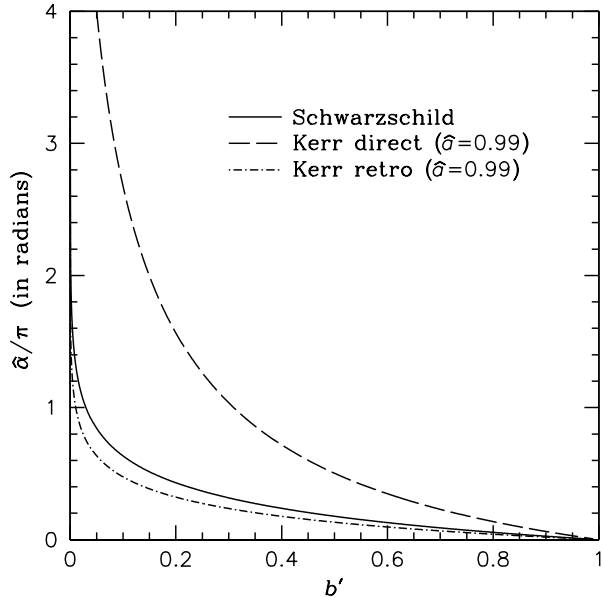}
\caption{\small \sl Exact bending angle plotted as a function of the normalized impact parameter.  The strong
deflection regime is towards the left of this plot as $b'$ goes to zero, and the weak deflection
regime is as $b'\rightarrow \infty$.  The Schwarzschild bending angle
is shown for reference.} 
\label{ExactKerrp99} 
\end{center} 
\end{figure}

We now have an explicit expression for the bending angle (\ref{ExactKAlpha}) 
via equations (\ref{bigcoeff})-(\ref{nplusminus}) in terms of $b'$, $a$ and $\mdot$, the
remaining independent quantities.  In Figures \ref{ExactKerrp5} and \ref{ExactKerrp99}, we have
plotted the exact bending angle as a function of $b'$ for $\hat{a}=0.5$ and $\hat{a}=0.99$.  Recall once again 
that the critical impact parameter depends on the spin parameter as well as on whether
the light ray is undergoing direct or retrograde motion.  As can be seen clearly 
from the bending angle plots, the bending angle is greater
than the Schwarzschild value for direct orbits, and {\em smaller} for retro orbits.  In addition for the 
higher spin ($\hat{a}=0.99$) case the effect is much more pronounced.  Plots for other values of
$\hat{a}$ look similar to this, and in the limit $\hat{a}=0$, all three curves merge as expected.

One of the first remarkable features is the suppression of the bending angle on the retro side compared to the 
Schwarzschild case.  On the direct side, light rays are bent more because of they are being
swirled ``downstream" with the spin, while motion of light rays ``upstream" results in {\em smaller} bending angles. In both
the direct and retro motion, as in the Schwarzschild case, the bending angle exceeds $2\pi$, resulting in
multiple loops and the formation of relativistic images.  As a direct consequence of the difference between the
direct and the retro side, as seen by an observer, these images will be offset to different amounts on either side.  

\begin{figure}[htp] 
\begin{center} 
\includegraphics[width=3.3in]{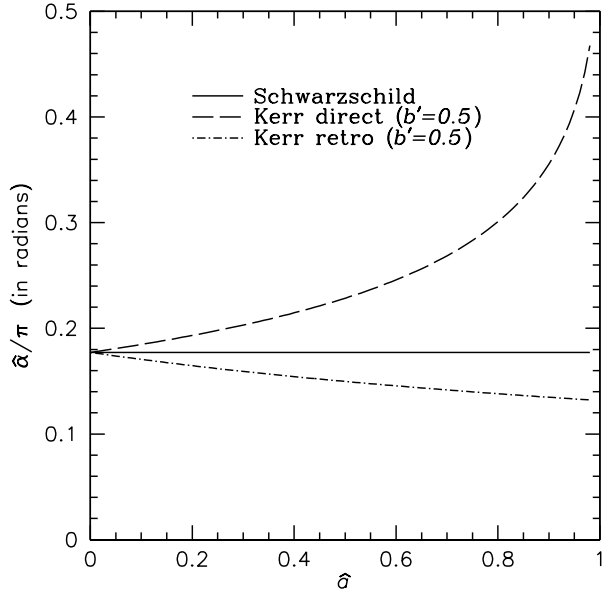}
\caption{\small \sl Bending angle plotted as a function of spin for $b^\prime=0.5$.  The Schwarzschild case is shown
for reference.} 
\label{AnglevSpin} 
\end{center} 
\end{figure} 

A plot of the bending angle divided by $\pi$ as a function of $\hat{a}$ for $b'=0.5$ is shown in 
Figure \ref{AnglevSpin} to illustrate the dependence on the spin.  The solid line
in this plot is the zero-spin Schwarzschild case shown for reference.

\section{Conclusions}

We have presented a detailed calculation of the exact bending angle formula for a light ray in the equatorial plane of
a Kerr black hole.  The frame-dragging in the Kerr geometry has a different effect on the light ray 
depending on whether it is traversing with or against the spin direction.  The extent to which
it is deflected from its original path is much higher for direct orbits than for retrograde orbits.  As a result, relativistic images
on the direct side shifts inward as $a$ increases, and outward on the retrograde side \cite{iyerhansen2}.
Higher order relativistic images can be studied by setting $\alphahat=2n\pi$ and using the solution
in the lens equation.  In order to demonstrate this, however, we first need to determine the
series expansions of the bending angle in both the strong and weak deflection limit.  We present 
these two series expansions in a companion paper \cite{iyerhansen2}.  A direct application of this
to the case when the source, lens and observer are perfectly lined up is also presented in \cite{iyerhansen2}.  

Analytical solutions and results derived from them are crucial to our understanding of lensing images in general, and
the Kerr geometry in particular.  Even with our results here being exact, we have only considered the equatorial Kerr 
case.  Our goal is to work with the full off-plane null geodesics in a similar manner to derive analytical 
solutions for predicting the behavior of a light ray as it spirals its way around the Kerr black hole before emerging towards the observer.

\vspace{0.2in}

\begin{acknowledgments}

S. V. I. thanks Arlie O. Petters for numerous helpful discussions.  E. C. H. was 
funded by the Dr. Jerry D. Reber Student/Faculty Research Fund at SUNY 
Geneseo.  The authors thank Kevin Cassidy for his support of this research fund.

\end{acknowledgments}

\end{document}